\documentclass{emulateapj}

\def\Chandra{${\it Chandra}$}

\newcommand{\Msun}{\ifmmode {M_{\odot}}\else${M_{\odot}}$\fi}
\newcommand{\Rsun}{\ifmmode {R_{\odot}}\else${R_{\odot}}$\fi}

\usepackage{graphicx}

\newcommand{\lsim }{{\lower0.8ex\hbox{$\buildrel <\over\sim$}}}
\newcommand{\gsim }{{\lower0.8ex\hbox{$\buildrel >\over\sim$}}}

\shorttitle{Ultracompact X-ray Binaries}
\shortauthors{Heinke et al.}

\begin{document}
\title{Galactic Ultracompact X-ray Binaries: Disk Stability and Evolution}
\author{C.~O. Heinke\altaffilmark{1}, N.~Ivanova, M.~C. Engel, K. Pavlovskii, G.~R. Sivakoff, T.~F. Cartwright\altaffilmark{2}, J.~C. Gladstone}
\affil{Physics Dept., 4-183 CCIS, Univ. of Alberta, Edmonton AB, T6G 2E1, Canada}
\altaffiltext{1}{Ingenuity New Faculty; heinke@ualberta.ca}
\altaffiltext{2}{International Space University, 1 rue Jean-Dominique Cassini, 67400 Illkirch-Graffenstaden, France}

\begin{abstract} 
We study the mass transfer rates and disk stability conditions of ultracompact X-ray binaries (UCXBs) using empirical time-averaged X-ray luminosities from Paper I (Cartwright et al. 2013) and compiled information from the literature.  The majority of UCXBs are consistent with evolutionary tracks for white dwarf donors. Three UCXBs with orbital periods longer than 40 minutes have mass transfer rates above $10^{-10}$ \Msun/year, inconsistent with white dwarf donor tracks.  We show that if helium star donors can retain their initial high entropy, they can explain the observed mass transfer rates of these UCXBs.

Several UCXBs show persistent luminosities apparently below the disk instability limit for irradiated He accretion disks.  We point out that a predominantly C and/or O disk (as observed in the optical spectra of several) lowers the disk instability limit, explaining this disagreement.    
The orbital period and low time-averaged mass transfer rate of 2S 0918-549 provide evidence that the donor star is a low-entropy C/O white dwarf, consistent with optical spectra. 

 We combine existing information to constrain the masses of the donors in 4U 1916-053 (0.064$\pm0.010$ \Msun) and 4U 1626-67 ($<$0.036 \Msun\ for a 1.4 \Msun\ neutron star).  
We show that 4U 1626-67 is indeed persistent, and not undergoing a transient outburst, leaving He star models as the best explanation for the donor.  

\end{abstract}



\maketitle

\section{Introduction}\label{s:intro}

Ultra-compact X-ray binaries (UCXBs) contain a neutron star (or black hole; black hole systems have not been confirmed yet) accretor and a compact donor star, with an orbital period $P_{orb}<80$ minutes.  The donors must be hydrogen-deficient, partially or fully degenerate stars \citep[e.g.][]{Rappaport82,Deloye03}.
UCXBs are preferentially produced in globular clusters (GCs), likely due to enhanced formation rates in such regions due to close dynamical interactions \citep{Verbunt87b,Deutsch00,Ivanova05,Ivanova10}. 

Three major scenarios for the nature of UCXB donors have been extensively discussed \citep[e.g.][]{Nelemans10}; we briefly review them below.  
A binary of a low-mass white dwarf (WD) and neutron star (NS) will lose angular momentum by gravitational radiation, forcing the WD to eventually begin transferring mass to the NS \citep{Pringle75}; we call this the WD evolution scenario.
Mass transfer starting as the donor ascends the subgiant branch can lead to decreasing periods, down to below an hour, as the degenerate, hydrogen-poor core is exposed \citep{Nelson86}; we call this the evolved main-sequence evolution scenario.  Finally, the donor star may be a helium star, burning helium in its core, at the time of contact \citep{Savonije86}, known as the He star evolution scenario.  Each scenario predicts different mass-transfer rates (due principally to the different entropies of the donor star) and different donor chemical compositions (see \citealt{Nelemans10b} for a review).  Thermonuclear X-ray bursts on NSs exhibit different characteristics depending on the nature of the fuel being burned (hydrogen or helium), and can therefore help constrain the composition of the accreted fuel \citep[e.g.][]{Cumming03,Galloway08}.  

UCXB systems can be roughly categorized as persistent (over the $\sim$decades we have been observing them) or transient.  Transient UCXBs spend the majority of the time in a quiescent state with little or no accretion, punctuated by occasional outbursts. Whether a UCXB will be persistent or transient depends upon the mass-transfer rate ($\dot{M}$) of the system. A high $\dot{M}$, and the resulting heating of the accretion disk through friction and X-ray irradiation, maintains the accretion disk in an ionized state with a high viscosity, which allows continued mass flow through the disk  \citep{Osaki74,White84,Lasota01}. If $\dot{M}$ from the companion is below some critical rate $\dot{M}_{crit}$, the mass transfer is unable to keep the entire accretion disk ionized.  The viscosity decreases, stopping mass flow through the disk until enough mass builds up to re-ionize the disk; this leads to transient behavior.  

The value of $\dot{M}_{crit}$ depends on the orbital period, as smaller disks require smaller $\dot{M}$ to maintain the entire disk in an ionized state \citep{Smak83}. It also depends upon the chemical composition of the disk, as lower-ionization-potential atoms allow faster ionization \citep{Menou02}, and upon the effects of irradiation of the disk, which keeps it ionized to lower mass-transfer rates \citep{Dubus99}.  Dubus et al. stressed that the numerical calculation of the effects of irradiation is still significantly uncertain.  \citet{Lasota08} noted that several persistent UCXBs appeared to be below the critical mass-transfer rate for stability of He disks, and suggested that the donors may contain some hydrogen, which would lower the critical mass-transfer rate.

\citet{Juett01} presented evidence that five X-ray binaries have unusual O/Ne ratios in their X-ray absorption spectra, suggesting that these donors were originally C/O or O/Ne/Mg WDs (though the Ne in many cases may be interstellar, \citealt{Juett05}, \citealt{Krauss07}). 4U 1626-67 shows clear evidence of C, O and Ne in X-ray \citep{Schulz01} and ultraviolet \citep{Homer02} spectroscopy, without evidence of helium.
\citet{Nelemans04,Nelemans06} and \citet{Werner06} presented evidence from optical spectroscopy that two  UCXBs (4U 1626-67, 4U 0614+09) clearly lack hydrogen and helium lines but exhibit carbon and oxygen lines, indicating that these systems have C/O WD donors. \citet{Nelemans06} also showed that 4U 1543-624 \& 2S 0918-549 show optical spectra similar to 4U 0614+09.  \citet{Dieball05} presented evidence from ultraviolet photometry requiring carbon in the disk of M15 X-2, and probably helium as well.  Broad oxygen emission lines have been identified in high-resolution X-ray spectra of 4U 1543-624 and 4U 0614+091 \citep{Madej10,Schulz10,Madej11}.  Thus C/O donors are likely common in UCXBs, and the mass-transfer stability behavior of C/O disks is likely to be important.  

This apparent lack of helium poses problems for our understanding of the observed X-ray bursts, as their properties indicate the presence of substantial amounts of helium \citep[see, e.g.][]{Juett03,intZand07,Kuulkers10}.   Spallation of heavy nuclei by accreting material to produce helium \citep{Bildsten92} has been repeatedly suggested as a possible solution, but suffers two well-recognized problems: the spallation requires infalling material of very different A/Z values, which is hard to understand in predominantly C/O accretion disks \citep{intZand05b}; and such spallation would also produce sufficient hydrogen to alter the characteristics of X-ray bursts, which in most UCXBs show no evidence of H \citep{Cumming03,Galloway08,Galloway10}.  

The overabundance of C/O disks among UCXBs has been widely discussed \citep[e.g.][]{Nelemans10b}, but is not understood, since standard population syntheses tend to produce far more He WD systems than C/O WD systems \citep{Nelemans10b,Belczynski04}.  Hybrid WDs of mass $<$0.45 \Msun are thought to be needed to explain the C/O disks \citep{Yungelson02}. The reason is that high-mass donor WDs will produce extremely high (super-Eddington) mass transfer rates when they make contact at small periods, and the accretor may be unable to expel the accreting material for $>$0.45 \Msun donor WDs.  It is unclear whether these hybrid WDs can also contain sufficient helium to produce X-ray bursts, without substantial fine-tuning of the model.

In Paper I \citep{Cartwright12}, we compiled a list of certain UCXBs, and used {\it Rossi X-ray Timing Explorer (RXTE)} bulge scan observations \citep{Swank01} and {\it Monitor of All-Sky X-ray Image (MAXI)} monitoring lightcurves \citep{Sugizaki11}, supplemented with \Chandra\ observations, {\it RXTE All-Sky Monitor (RXTE/ASM)} data, and literature reports, to construct histograms of the luminosities of these UCXBs.  With these measurements, we constructed an empirical luminosity function for galactic UCXBs.  Here we use the time-averaged luminosities from Paper I to interpret the behavior of individual UCXBs where the period is known (or suggested), reviewing a wide body of literature to identify critical information. We give relevant information for these UCXBs in Table 1.

\begin{table*}[h!]
\begin{tabular}{ c@{}c@{ }c@{}c@{  }c@{}c@{}c@{}c }
  Source & Location & Distance & Period & $N_H$  & Average mass transfer  & Spectral & Bursts? \\
    & & (kpc) & (minutes) & ($10^{21}$ cm$^{-2}$) & \Msun/year & data & (nature) \\
  \hline  %
  \multicolumn{8}{c}{Persistent systems} \\
  \hline
 4U 1728-34 & Bulge& $5.2\pm0.8^{a}$ & $10.8?^{a}$ & $22.9^{a}$ & $2.0\pm1.2\times10^{-9}$  & - & Yes; He$^{a}$ \\
  4U 1820-303 & GC& $7.9\pm0.4^{b}$ & $11^c$ &  1.6$^d$  & $1.2\pm0.6\times10^{-8}$ & - & Yes; He$^{\theta}$  \\
  4U 0513-40 & GC & $12.1\pm0.6^b$ & $17^e$  &  0.26$^d$  & $1.2\pm0.6\times10^{-9}$ & - & Yes \\
  2S 0918-549 & Field & $5.4\pm0.8^f$ & $17.4^g$ & 3.0$^h$ & $2.6\pm1.5\times10^{-10}$ & C/O?$^{\kappa,\lambda}$  & Yes; He$^f$ \\
  4U 1543-624 & Field & $7.0?^i$ & $18.2^i$ & 3.5$^h$ & $1.3^{+1.8}_{-1.2}\times10^{-9}$ & C/O?$^{\kappa}$;O$^{\mu}$ & No \\
 4U 1850-087 & GC & $6.9\pm0.3^b$ & $20.6^j$ & 3.9$^d$ & $2.2\pm1.1\times10^{-10}$ & - & Yes \\
  M15 X-2 & GC & $10.4\pm0.5^b$ & $22.6^k$ & 0.67$^l$ & $3.8\pm1.9\times10^{-10}$ & C,He$^{k}$ & Yes \\
 4U 1626-67 & Field & $8^{+5\ }_{-3}$ $^m$ & $42^m$ & 1.4$^n$ & $8^{+14}_{-6}\times10^{-10}$ &  C,O,Ne$^{\xi,\pi,\sigma,\lambda}$ & No \\
 4U 1916-053 & Field & $9.3\pm1.4^o$ & $50^p$ &  3.2$^q$ & $6.3\pm3.7\times10^{-10}$ & He,N$^{\lambda}$ & Yes \\
 4U 0614+091 & Field & $3.2\pm0.5^r$ & $51?^s$ & 3.0$^t$ & $3.9\pm2.3\times10^{-10}$ & C/O$^{\kappa,\lambda,\sigma}$;O$^{\tau,\phi}$ & Yes \\
 \hline %
  \multicolumn{8}{c}{Transient systems} \\
  \hline
  XTE J1807-294 & Bulge & $8^{+4}_{-3.3}$ $^x$ & $40.1^y$ & $5.6^{z}$ & $<1.5^{+1.9}_{-1.2}\times10^{-11}$ & - & No\\
 XTE J1751-305 & Bulge & $8^{+0.5\ }_{-1.3}$ $^{\alpha}$ & $42^{\beta}$ & $9.8^{\gamma}$ & $5.1^{+2.6}_{-2.9}\times10^{-12}$ & - & No \\
  XTE J0929-314 &  Field & $8^{+7\ }_{-3}$ $^x$ &  43.6$^{\delta}$ & 0.76$^{\epsilon}$ & $<9.7^{+25}_{-7.7}\times10^{-12}$ & C/O?$^{\lambda}$ & No \\
 Swift J1756.9-2508 & Bulge & $8\pm4^{\zeta}$ & 54.7$^{\zeta}$ &  54$^{\zeta}$ & $1.7^{+2.3}_{-1.5}\times10^{-11}$ & - & No \\
 NGC 6440 X-2 & GC & 8.5$\pm0.4^b$ & 57.3$^{\eta}$ & 5.9$^b$ & $1.0^{+0.5}_{-0.5}\times10^{-12}$ & - & No \\
 \end{tabular}
\caption{UCXBs with known or suggested periods, with best estimates of their distance and $N_H$, our estimate of their mass transfer rate using the monitoring in Paper 1, and literature information on identification of spectra and properties of thermonuclear bursts (He means the accreted matter lacks H). Location in the Galactic field, (direction of the) bulge, or in a Globular Cluster (GC) is also specified.
Distance errors are ranges from indirect estimates; 15\% errors on bursts \citep{Kuulkers03}; 5\% errors on GC distances.  Periods and compositions  supported by only weak evidence have '?'s. Spectral data from optical, X-ray, or UV spectroscopy.  
 References: $^{a}$\citet{Galloway10}; $^{b}$\citet{Harris10}; $^{c}$\citet{Stella87}; $^{d}$\citet{Sidoli01}; $^{e}$\citet{Zurek09}; $^{f}$\citet{intZand05b}; $^{g}$\citet{Zhong11}; $^{h}$\citet{Juett03b}; $^{i}$\citet{Wang04}, distance estimate assumes $\dot{M}$ driven by GR; $^{j}$\citet{Homer96}; $^k$\citet{Dieball05}; $^l$\citet{White01}; $^m$\citet{Chakrabarty98}; $^n$\citet{Krauss07}; $^o$\citet{Yoshida93}; $^p$\citet{Walter82}; $^q$\citet{Church98}; $^r$\citet{Brandt92}; $^s$\citet{Shahbaz08}; $^t$\citet{Piraino99}; 
 $^x$\citet{Galloway06b}; $^y$\citet{Markwardt03}; $^z$\citet{Falanga05}; $^{\alpha}$\citet{Papitto08}; $^{\beta}$\citet{Markwardt02}; $^{\gamma}$\citet{Miller03}; $^{\delta}$\citet{Galloway02}; $^{\epsilon}$\citet{Juett03}; $^{\zeta}$\citet{Krimm07}; $^{\eta}$\citet{Altamirano10}; $^{\theta}$ \citet{Cumming03};  $^{\kappa}$\citet{Nelemans04}; $^{\lambda}$\citet{Nelemans06}; $^{\mu}$\citet{Madej11}; $^{\xi}$\citet{Schulz01}; $^{\pi}$\citet{Homer02}; $^{\sigma}$\citet{Werner06}; $^{\tau}$\citet{Schulz10}; $^{\phi}$\citet{Madej10}.  
}
\label{all_sources}
 \end{table*}

\section{Time-averaged luminosities vs. period}

\begin{figure}
\figurenum{1}
\includegraphics[scale=0.5]{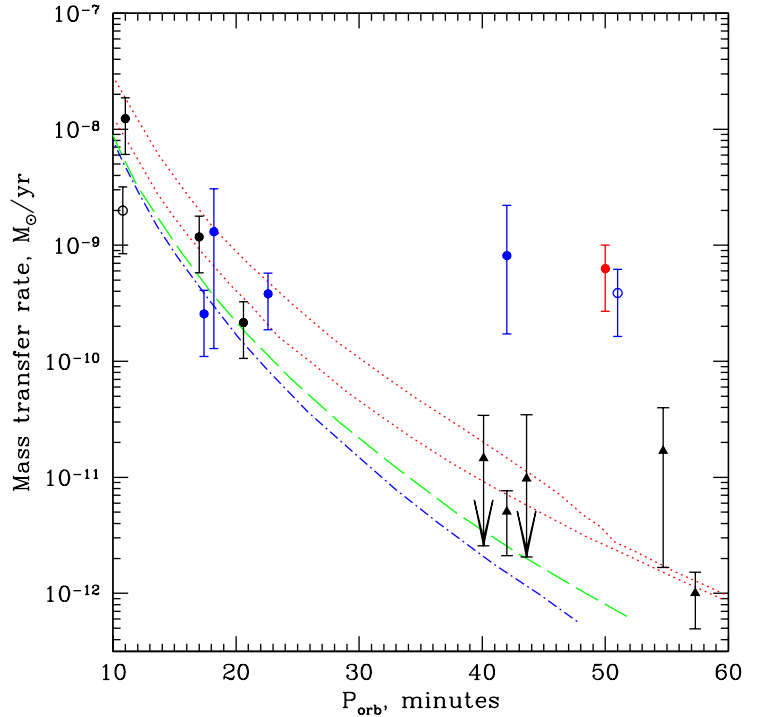}
\caption{Observed periods and $\dot{M}$ (from long-term X-ray lightcurves) of UCXBs, with persistent systems as dots, transients as triangles (with $\dot{M}$ upper limits for transient systems with only one recorded outburst). $\dot{M}$ values were calculated from the information in \citet{Cartwright12}. Errors correspond to a 25\% error in the bolometric correction factor (2.9), combined in quadrature with errors in distance. Symbol colors indicate spectroscopic information on disk composition; red for helium, blue for carbon and/or oxygen, black for no information. 4U 1728-34 and 4U 0614+091 are marked with open symbols, to indicate that the suggested orbital periods may not be correct. 
Theoretical evolution tracks for WD UCXBs from C. Deloye and \citet{Lasota08} are plotted as red dotted lines, for an initially 0.325 $\Msun$ secondary, a 1.4 $\Msun$ NS, and degeneracy parameters $\psi$ of 3.0 (lower curve; lower entropy donor) and 1.5 (upper curve; higher entropy donor).  Additional tracks from \citet{Deloye03} give low-entropy tracks for C (green long-dashed) and O (blue dash-dotted) donors. Note that the majority of UCXBs are consistent with WD UCXB evolutionary tracks, apart from three systems with high mass transfer at long periods, and possibly 4U 1728-34 at 10.8 minutes.
}
\label{degtrack}
\end{figure} 

We construct a version of \citet{Lasota08}'s plot of UCXB $\dot{M}$ vs. orbital period in Fig. \ref{degtrack} using our time-averaged X-ray luminosity calculations (Table 1; see Paper I for details on the derivation) and a bolometric correction of 2.9 \citep{intZand07}. We estimate a $\sim$ 25\% error associated with the bolometric correction factor and use distance errors as given in Table 1  (for 4U 1543-624, we estimate a distance uncertainty of $\sim$ 50\%, see Paper I). These errors are combined in quadrature to produce upper and lower error estimates for $\dot{M}$ (Table 1, Fig. 1). In calculating $\dot{M}$, we assume a 1.4 $\Msun$ NS with an 11.5 km radius \citep[e.g.][]{Steiner12}.  (Predicted model mass transfer rates vary linearly with the NS mass, while luminosity-inferred mass transfer rates vary inversely with the NS mass.  Nevertheless, the errors introduced by these choices are generally smaller than the other uncertainties.)  We differ from many previous works in regarding 4U 1626-67 as a persistent, rather than transient, source, considering its short orbital period (and thus small disk), 40 years of continuous activity (vs. $\sim$month-long outbursts of transient UCXBs of similar periods), and currently increasing $L_X$ (see below).  $\dot{M}$ values for 4 of the transient sources are very similar to previous literature estimates by \citet{Galloway06b} for XTE J1807-294 and XTE J0929-314, \citet{Heinke09a} for XTE J1751-305, and \citet{Heinke10} for NGC 6440 X-2.  Our range for Swift J1756.9-2508 (rather wide, due principally to its distance uncertainty) is consistent with that of \citet{Patruno10}, although it still includes the lower estimate of \citet{Krimm07}.  The general features of this plot are similar to the calculations by \citet{vanHaaften12}, though we include an additional source (NGC 6440 X-2) and obtain smaller error bars for our fainter sources.

The timescales over which these average mass transfer rates are calculated vary from 2.7 years (for the MAXI sources), to 15 years for the two sources where RXTE ASM data was used, and is typically 12 years for the PCA bulge scan sources.  The length of the timescales used limits our accuracy in determining the mass-transfer rate, as systems may go through cycles of enhanced mass transfer on timescales longer than our datasets \citep{Kotze10}.  Evidence of such variations include the well-studied decline (and now rise, by a similar factor of $\sim$5) of 4U 1626-67's mass transfer \citep{Chakrabarty97,CameroA10, Jain10}, and the limited series of outbursts from NGC 6440 X-2 \citep{Heinke10}.  The nature of such cycles is not clear (especially for degenerate donors), but the possibility of such behavior must be considered as a caveat when interpreting our results.

\subsection{Evolutionary tracks}

We first consider the evolution of WD UCXBs, as discussed in detail by \citet{Deloye03} and \citet{vanHaaften12b}.
We plot, in Fig.\ 1, tracks for adiabatic UCXB evolution from C. Deloye, as shown in \citet{Lasota08}, for low-entropy or high-entropy He donors (dotted lines), and tracks for low-entropy C or O donors (dashed lines) from \citet{Deloye03}.  These tracks consider that WDs are not completely degenerate at the start of the mass transfer, and hence have some final entropy in the center, providing some range in possible mass transfer rates for the same period \citep{Deloye03}.  This effect decays for long periods, as the donor has time to thermally relax, explaining the return of the high-entropy track towards the low-entropy track \citep{Deloye07}.

These tracks reasonably describe most of our UCXBs, apart from 4U 1728-34 (which does not have strong evidence yet for the suggested period) and three interesting long-period objects with high mass transfer rates (the period of 4U 0614+091 is also not certain), which we exclude for the moment. Using \citet{Deloye03}'s calculations of UCXB number distributions for assumed $n_{ad}$=-0.2 (appropriate for helium donors), we expect 6 times more He UCXBs between 15 and 25 minutes as between 5 and 15, consistent with the relative numbers (4U 1820-30 vs. the five systems in the later bracket).  Slightly larger numbers of longer-period systems are expected for C or O donors.  We expect 29 times more He UCXBs on these tracks with 25$<$P$<$60 minutes as 5$<$P$<$25 minutes, giving a total of $\sim$175 UCXBs beyond 25 minutes, many more than the five known.  The unseen long-period systems might be persistent, be transient, or have stopped mass transfer.  

Persistent systems beyond 25 minutes, with expected $L_X<10^{36}$ ergs/s, would be hard to detect and characterize, though they are unlikely to remain persistent at significantly longer periods (see the next section).  Transient systems may not be detected because they show rare outbursts (e.g. XTE J0929-314) or more frequent outbursts that are too faint to be noticed in most surveys (e.g. NGC 6440 X-2).  Unfortunately, there does not seem to be any pattern to the changes in either outburst lengths or recurrence times (individually), as the mass transfer rate decreases.

Alternatively, many of these systems may have stopped accreting mass, allowing the NSs to become radio millisecond pulsars.  It is difficult to identify the pulsar descendants of UCXBs, since most short-orbital-period millisecond pulsars have orbital periods longer than predicted \citep{Deloye08b,Ivanova08}.  This has led to the suggestions that the donor stars are strongly heated and inflated, leading to longer periods \citep{Rasio00, Bailes11} and/or complete donor destruction \citep{Bildsten02,vanHaaften12b}, or to the possibility that many of these NSs are spun down as mass transfer decreases, and never become pulsars (\citealt{Jeffrey86,Deloye08}; but cf. \citealt{Tauris12}  which argues against this).  

\citet{vanHaaften12c} argue for enhanced UCXB angular momentum loss by donor wind mass loss to explain the companion of the millisecond pulsar PSR J1719-1438, which is a low-mass degenerate star in a 2.2 hour orbit \citep{Bailes11}.  This scenario offers the appeal of simultaneously explaining the long period of this system (too long for standard evolution to produce during a Hubble time), and perhaps of explaining the high mass transfer rates of several longer-period UCXBs \citep{vanHaaften12}.  \citet{vanHaaften12c} note that a similar wind mass loss is suggested to explain the orbital period derivative of SAX J1808.4-3658 \citep[e.g.][]{Burderi09}.  This scenario, however, suffers some difficulties.  The orbital period derivative of SAX J1808 is accelerating \citep{Patruno12}, indicating that the orbital period evolution is probably driven by exchange of angular momentum between the donor star and the orbit, as seen in many other binaries \citep[e.g. PSR 1957+20,][]{Arzoumanian94}.  The suggested mass loss rates may also be hard to achieve at this stage of their evolution.  ``Black widow'' radio-eclipsing millisecond pulsars such as PSR 1957+20 show ablative winds of only $\sim10^{-13}-10^{-14}$ \Msun/year \citep{Fruchter92,Eichler95}.  (These companions are irradiated by pulsar winds rather than X-rays, which will alter the physics of heating, but they intercept similar energy fluxes from the primary.)  However, those lower-density black widow companions have surface gravities that are factors of $\sim 10^4$ lower than similar-mass WDs, using the mass and radius values for the PSR 1957+20 companion, $M$=0.034 \Msun\ and $R\geq$0.25 \Rsun, from \citep{vanKerkwijk11}. (As we see below, this is roughly the appropriate mass for two high-mass-transfer long-period UCXBs.)  
A final concern is that it is difficult to understand how initially similar UCXBs with WD companions can evolve to the dramatically different mass loss rates seen among UCXBs with periods beyond 40 minutes in our sample.
Thus, we are skeptical that donor winds of the required magnitude are driven from UCXB  companions, but the potential of this mechanism to solve several problems strongly motivates further study of this possibility.  


\subsection{High-$\dot{M}$ systems}

There are two groups of UCXBs with periods of 40-60 minutes, well-separated in their time-averaged mass transfer rates; transient sources with low rates, $\la 10^{-11} M_\odot$ per yr, vs. persistent sources with average rates $\sim$100 times higher for the same period range (4U 1626-67, 4U 1916-053, and 4U 0614+091).
These persistent sources require an explanation other than simple WD evolution.  

One alternative is to invoke an angular momentum loss mechanism that is stronger than gravitational wave radiation. 
Donor wind mass loss (\citealt{vanHaaften12}; see discussion above) is one possibility. 
 Alternatively, if an accretor does not accept all the donor's material, a circumbinary disk (CBD) could form. A CBD, as it rotates slower than the binary orbits, provides a tidal torque on the binary, removing its orbital angular momentum  \citep[see][for more details on a simple CBD model]{Spruit01}. The strength of that tidal torque depends on the physics of the CBD, mainly its viscosity and the scale height, as well as on what fraction of the donor's mass loss ends up in the CBD, denoted as $\delta$. The physics of the CBD can be  further simplified, as for a standard $\alpha-$viscosity disk the loss of angular momentum takes a simple form \citep{Shao12}:
 
\begin{equation}
\dot J_{\rm CBD} = A (GM)^{2/3} \delta \dot M_{donor} t^{1/3}
\end{equation}
\noindent Here $A=(3\alpha \beta^2/4)^{1/3}$, where $\alpha$ is the viscosity parameter and $\beta$  is the ratio of the scale height of the disk to its radius; with values for a standard disk $\alpha=0.01$ and $\beta=0.03$, $A\approx0.02$. $t$ is time after the start of the mass transfer. Studies performed for cataclysmic variables suggested that $\delta\ll 1$, and its value is of order $10^{-4}$ to $10^{-3}$ \citep{Taam03}.  Further analysis showed that for a standard CBD in a binary system with an orbital period of about 1 hour, $\delta\sim 6\times 10^{-4} t_9^{-1/3}$ (where $t_9=t/10^9$ years; \citealt{Shao12}).

\begin{figure}
\figurenum{2}
\includegraphics[scale=0.7]{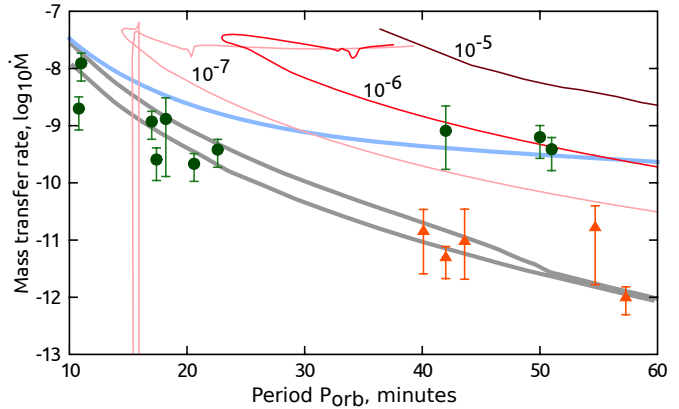}
\caption{Various evolutionary tracks; standard He WD UCXB evolution (low- and high-entropy cases from C. Deloye), black; 
a UCXB with an extreme circumbinary disk (CBD), cyan; and our He star evolutionary tracks, at initial mass loss rates of $10^{-7}$ \Msun/year (pink; note that there is a short period where mass transfer stops), $10^{-6}$ \Msun/year (red), and $10^{-5}$ \Msun/year (purple).  Persistent and transient UCXBs are labeled with green circles and orange triangles, respectively.  Note that either the CBD or He star evolutionary tracks can explain the persistent long-period UCXBs.
}
\label{fig:mdots}
\end{figure}

Using this formalism, a WD donor can drive a mass transfer rate of $2\times 10^{-10} M_\odot$ per yr at a period $>40$ minutes (note this is the minimum mass transfer rate in the second group), if the CBD model has  $\delta \ga 0.0025$. This is several times larger than it should be for this evolutionary stage, typically $\delta\sim 3\times 10^{-4}$.  See Fig. \ref{fig:mdots} where we show an example WD track with a very strong CBD included.

A second option begins with an initially slightly evolved main sequence donor with a helium-rich core, where the orbit shrinks due to magnetic braking, and can reach ultracompact periods \citep{Nelson86,Podsiadlowski02}.  Such an evolutionary sequence can produce mass transfer rates of $10^{-9}$ \Msun/year, and thus explain the second group of systems. This evolution requires rather finely tuned initial parameters to reach ultracompact orbits, producing very few  systems with periods below 1 hour, and almost none below 30 minutes \citep{vanderSluys05}.  This evolutionary sequence may leave some hydrogen in the core, a clear observable difference with the other sequences \citep{Nelemans10b}.

\begin{figure}
\figurenum{3}
\includegraphics[scale=0.7]{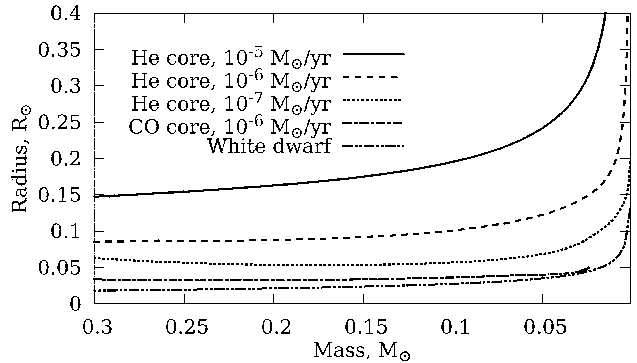}
\caption{Evolution of mass and radius for He stars at different initial mass loss rates from our {\tt MESA} calculations, including a He star stripped down to its CO core, vs. the evolution of a WD (\citealt{Tout97}, their eq. 17, a rough approximation for either He or CO WDs).  Note that the radii for He stars are substantially larger than WDs of the same mass.
}
\label{fig:He_star_MR}
\end{figure}

\begin{figure}
\figurenum{4}
\includegraphics[scale=0.7]{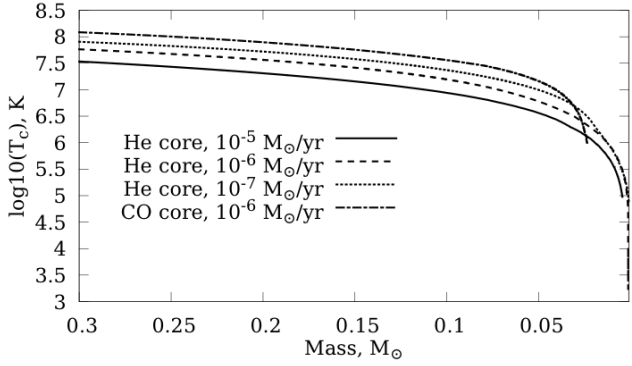}
\caption{Temperature evolution with time of He stars at different initial mass loss rates, including a He star stripped down to its CO core, from our {\tt MESA} calculations.  Note that He stars can retain high central temperatures down to very low masses.
}
\label{fig:He_star_T}
\end{figure}

\begin{figure}
\figurenum{5}
\includegraphics[scale=0.7]{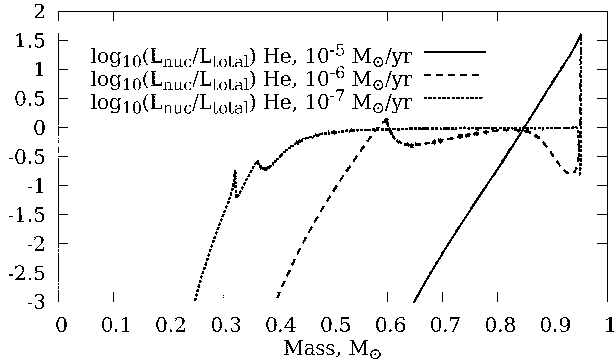}
\caption{Evolution of the nuclear burning luminosity as a fraction of the total luminosity for He stars at different initial mass loss rates from our {\tt MESA} calculations.  Note that the nuclear burning turns off more quickly for higher mass-loss rates due to rapid donor expansion.  The large initial nuclear luminosity (which may be larger than the total emitted luminosity) is due to the just-completed common envelope stage.
}
\label{fig:He_star_nuc}
\end{figure}

\begin{figure}
\figurenum{6}
\includegraphics[scale=0.7]{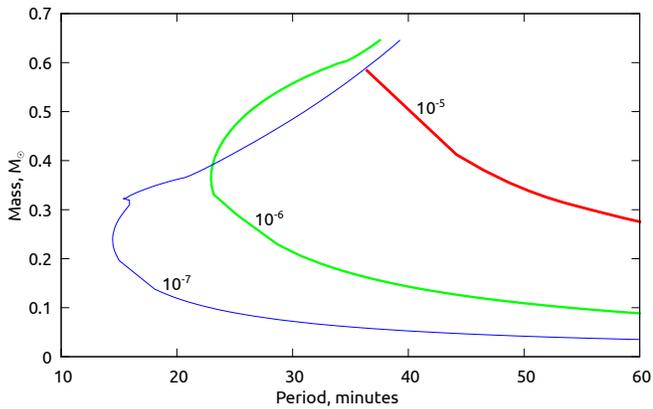}
\caption{Mass evolution vs. orbital period of He stars at different initial mass loss rates; tracks as in Fig. 2. These tracks are consistent in masses, orbital periods, and mass transfer rates (see Fig. 2) with the known information on 4U 1626-67 and 4U 1916-053 (\S 3.3, 3.4).
}
\label{fig:He_star_mass}
\end{figure}

For a third alternative, a He star can be produced by a common envelope event and inspiral via gravitational waves until it makes contact at short periods while still fusing He at its center (see \citealt{Yungelson08,Nelemans10b,vanHaaften12}). This avoids the fine-tuning difficulties with the evolved main-sequence star evolution. 
Naked He star donors generally have radii much larger than WDs of the same mass, where this radius is also a function of its final entropy (we demonstrate this dependence in Figs. \ref{fig:He_star_MR} and \ref{fig:He_star_T}, where we show radius and central temperature $T_{\rm c}$ evolution for a naked He core of a giant with initial mass of 5 $M_\odot$, evolved with different mass loss rates using the stellar code {\tt MESA}\footnote{MESA (Modules for Experiments in Stellar Astrophysics) is a collection of libraries for computational stellar astrophysics that is relied upon by a natively implemented one-dimensional stellar evolution code capable of modelling stars at a wide range of evolutionary stages \citep{Paxton11,Paxton13}.}).  Once a naked He star is formed -- through a common envelope event --
it also may start He burning in the core (note that whether it burns
or not depends on the naked He star mass). This burning may be fully
completed (in the sense that the core is fully converted to a carbon-oxygen
core) during the mass loss sequence, depending on the initial
post-common envelope binary separation, on how close to the giant tip
the donor was before the common envelope event, and how fast the He star
is losing mass (see Fig. \ref{fig:He_star_nuc}).
As a result of this burning, a very low-mass carbon-oxygen donor can
be formed, although in some cases He fusion is simply stopped by the 
rapid expansion of the donor and drop of its central temperature.  For example,
we note that very rapid mass loss leads to donor expansion and
cooling, hence burning is rapidly depleted and a C/O core might not form.
Continued He burning is more likely to provide an inflated donor,  vs.
a donor that had a composite He/C/O core before the mass transfer (see
Figs. \ref{fig:He_star_MR} and \ref{fig:He_star_T}). At some point, the 
He donor starts to expand with continued mass loss - note that this can happen due to various
reasons, e.g., due to the core's conversion into a C/O core, or to 
adiabatic expansion due to rapid mass loss. The point where the
nuclear burning turns off can also be roughly identified as where the He
star tracks begin to expand outwards to longer periods again (see Fig.
\ref{fig:He_star_mass}.

If a He star {\it retained} its entropy in the center due to either faster (in the past) mass transfer, or due to nuclear burning, tidal heating, or ongoing irradiation, it can provide the observed mass transfer rates requiring only gravitational wave radiation  without invoking a CBD. 
We tested this situation by applying a fast mass loss rate to a He star model in {\tt MESA}, and then checking what mass transfer rate this star (which is out of thermal equilibrium) will have if it is in a binary evolving only under gravitational radiation (see Fig. \ref{fig:mdots}).   
Nondegenerate He star cores (appropriate for the intermediate-mass progenitors required) expand upon mass loss from the outer envelope  \citep{Deloye10,Ivanova11,Ivanova12}.  Thus, they can continue to drive mass transfer as the orbit expands, giving mass transfer rates up to $10^{-3}$ \Msun/year for some fraction of the core.  A fully self-consistent calculation of the evolution of He stars in binary systems has not yet been performed, and the stage where the core will stop expanding is not well established.  MESA does not include tidal heating or donor irradiation, which could increase the inferred mass loss rates.

 If the donor is an (inflated) He star evolving under gravitational wave radiation only, then mass transfer rates do not significantly exceed the Eddington rate during most of the mass loss evolution, while still transferring $\sim0.5$ \Msun\ (Fig. \ref{fig:He_star_mass}). If an accretor accepts most of the transferred mass below the Eddington limit, then the NS in such a binary could grow significantly more massive than a NS with a WD donor, potentially forming a NS with $M\ga 2 M_\odot$.  This is in contrast with WD UCXB evolution; cold WDs more massive than 0.08 \Msun, upon starting mass transfer, will exceed the Eddington limit \citep{Bildsten04}, so cannot efficiently transfer more than $\sim$0.1 \Msun\ to the NS.

\clearpage

\clearpage

He stars may show a wide range of surface abundances, depending on the initial post-common-envelope orbital period, which sets how long the star burns He before mass transfer stops fusion.  \citet{Nelemans10b} present extensive calculations of the evolution and abundances of He stars.  In the initial, rapid epoch of orbital shrinkage, the outer, unburnt helium layers are consumed (note that this epoch is very brief, $<10^7$ years, and thus difficult to observe).  After period minimum, the C/O fusion products are revealed, though some He is still available. \citet{Nelemans10b} show that wider initial orbital periods (e.g. 200 minutes) give primarily C and O chemical compositions at $P_{orb}>30$ minutes, with reduced He.  This matches the inferred compositions of 4U 1626-67 (substantial C, O, and Ne, \citealt{Schulz01,Werner06}) and 4U 0614+09 (C and O dominate \citealt{Nelemans04}; while X-ray bursts indicate the  presence of He without H,   \citealt{Kuulkers10}).  \citet{Schulz01} claimed an overabundance of Ne in local absorbing material around 4U 1626-67, which would be hard to explain with a He star (or any star hot enough to provide this mass transfer rate), since the Ne can only sink to the core in a cold WD.  However, the evidence for overabundant Ne in absorption 
 seems to have disappeared, leaving only the strong Ne X-ray emission lines as more ambiguous evidence for Ne's abundance \citet{Krauss07}.  

We note that the He star mechanism to create longer-period, high-mass-transfer systems cannot work in globular clusters, due to the relatively high masses of the inital donors ($>$2.3 \Msun) and short lifetimes of the systems \citep{Yungelson08}.  This is consistent with the lack of UCXB systems with unusually high mass transfer rates for their orbital period, like 4U 1626-67 or 4U 1916-053, in globular clusters.  The lack of the He star mechanism in globular clusters also can explain part of the difference in the distribution of orbital periods between globular cluster and field UCXBs (that field UCXBs have longer periods) which was noted by \citet{Zurek09}.  

\subsection{Critical mass-transfer rates of He vs. C/O accretion disks}

\citet{Deloye03}  and \citet{Lasota08} applied accretion disk stability calculations to understand the behaviour of UCXBs. \citet{Deloye03} calculated the evolution of WD donors evolving to lower mass transfer rates as they move outwards, reaching mass transfer rates prone to disk instability (for solar composition) as they reach orbital periods around 30 minutes and $\dot{M}_{crit}=3-6\times10^{-11}$ \Msun/yr. \citet{Lasota08} pointed out that irradiated pure He disks require higher $\dot{M}$ for stability, and thus become unstable earlier, at orbital periods around 20 minutes and $\dot{M}_{crit}=3-5\times10^{-10}$ \Msun/yr.  Lasota et al. (2008) also noted that this stability criterion lies above three known persistent systems, and suggested that this can be resolved by the donors retaining a small fraction of H, e.g. by the evolutionary models of \citet{Podsiadlowski02}. These models have serious  difficulties explaining the numbers of UCXBs at very short periods \citep{vanderSluys05}.  Here, we show that the possession of carbon and oxygen in the disks of most of these UCXBs solves the problem.

\begin{figure}
\figurenum{7}
\includegraphics[scale=0.43]{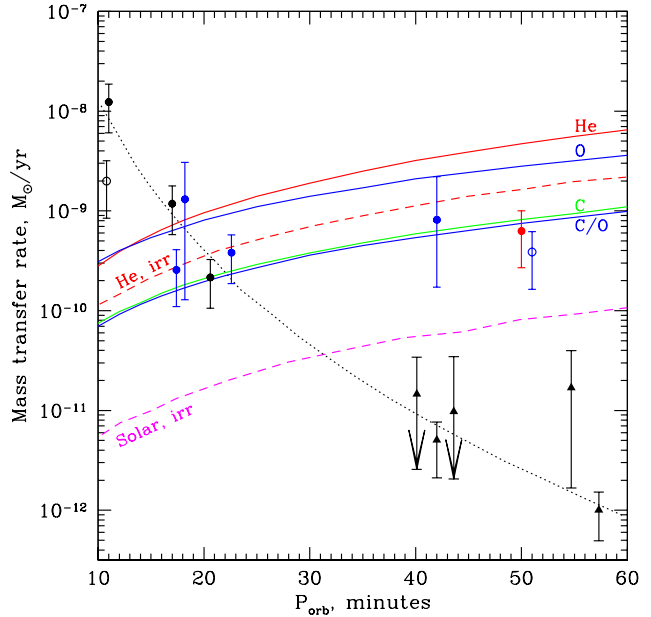}
\caption{Stability limits for accretion disks of specified compositions and a radius equal to 2/3 the accretor's Roche lobe radius are plotted.  Helium in red (solid line, for an unirradiated disk; dashed line, irradiated), oxygen in blue, carbon in green, C/O also in blue, and solar metallicity material (irradiated) in magenta.   Irradiated disk limits (dashed lines) from \citet{Lasota08}, others (solid lines) from \citet{Menou02}.  The lower-entropy helium WD evolution track \citep{Deloye03} is plotted, as in Fig. 1.  The meaning of the UCXB datapoints is as described in Fig. 1.  Note that the unirradiated disk stability limits for C/O disks are $\sim$5 times lower than unirradiated He disk stability limits, and that irradiation will lower them further, thus reasonably explaining the persistence of the persistent UCXBs.
}
\label{fig:stability}
\end{figure} 

We take stability curves for He, C, O, and C/O disks from \citet{Menou02}, assuming $M_{NS}=1.4$ $\Msun$  and $\alpha=0.1$.
The stability curves for C and C/O disks lie below most of the lowest-luminosity persistent sources (Fig. \ref{fig:stability}).  
 We also plot the irradiated disk $\dot{M}_{crit}$ lines from \citet{Lasota08} for He and solar composition.  Accurate $\dot{M}_{crit}$ calculations for irradiated C and/or C/O disks have not yet been done, but are clearly needed.  However, we might estimate from the drop in $\dot{M}_{crit}$ for irradiated vs. non-irradiated He disks that the $\dot{M}_{crit}$ for irradiated C/O disks will probably cross the Deloye helium UCXB tracks around $10^{-10}$ \Msun/yr, below all persistent UCXBs with known periods.  Four of the six persistent UCXBs at or below the irradiated He stability line (4U 0614+091, 4U 1626-67, 2S 0918-549, and M15 X-2) show strong C or O lines in their optical or UV spectra  \citep{Nelemans04,Nelemans06, Dieball05}. Thus, C/O disks seem appropriate for them, and indicate they should be persistent, as observed. One of the other two persistent UCXBs below the helium stability line is 4U 1916-053, which shows He and N lines, but no C or O  (the other, 4U 1850-087 in a globular cluster, has no spectral information). We suspect that the nitrogen (a typical product of CNO cycle burning) could also lower the stability line, though to our knowledge no calculations for such disks have been performed.

\section{Individual UCXBs}
Here we consider the detailed properties of a few individual persistent systems, using all available information to constrain their evolutionary history.  These five systems do not lie on the standard He WD UCXB tracks, though two of them have rather uncertain orbital periods (4U 1728-34 and 4U 0614+091).

\subsection{2S 0918-549}
One persistent UCXB with a well-determined period and distance, 2S 0918-549, has a mass-transfer rate significantly below the low-entropy helium UCXB track (Fig. 1).  Such a determination was suggested by \citet{Deloye03} as a method to securely identify a UCXB donor as C/O rather than He.  This determination is supported by \citep{Nelemans04}, who identify 2S 0918-549's optical spectrum as closely resembling the disk of 4U 0614+09, which shows only C and O lines without detectable H or He (though some He, perhaps 10\%, might remain). The mass-transfer stability requirement \citep{Yungelson02,Nelemans10b} suggests that the donor is a hybrid WD of initial mass $<$0.45 \Msun. However, the helium mantle of such objects is generally lost at very short periods, so the observation of (likely helium-powered) X-ray bursts \citep{intZand05} from this system is hard to explain.

\subsection{4U 1728-34}
4U 1728-34 has a suggested orbital period of 10.8 minutes \citep{Galloway10}, from Chandra data.  Its time-averaged luminosity indicates a mass-transfer rate a factor of 3-5 below the predictions of any UCXB evolutionary track for an 11-minute period.  This time-averaged luminosity is consistent with all X-ray measurements over the 30-year history of X-ray astronomy.  If the period is verified, we would not see any alternative to requiring substantial variations in its mass-transfer rate on timescales $>$ 30 years. 

\subsection{4U 1916-053}
 4U 1916-053 shows strong He and N lines in its optical spectra, indicating a predominantly He donor with CNO-processed material \citep{Nelemans06}.  4U 1916-053 shows evidence for precession of its accretion disk, by showing a ``superhump'' optical period 0.9\% longer than its true binary period \citep{Chou01,Retter02}.  An empirically calibrated relation between the mass ratio, $q=M_2/M_1$, and the fractional excess $\epsilon=(P_{sh}-P_{orb})/P_{orb}$ (where $P_{sh}$ is the [longer] superhump period), was shown for cataclysmic variables by \citet{Patterson05} (see also \citealt{Pearson06,Patterson01}), $\epsilon=0.18 q + 0.29 q^2$.  Assuming that this relation also works for X-ray binaries (justified by the few low-mass X-ray binaries considered in \citealt{Patterson01}), we find $q=0.046$, 
 which for a NS mass of $1.4\pm0.2$ \Msun\ (a range including the majority of well-measured NS masses), gives a companion mass of 0.064$\pm0.010$ \Msun.  Requiring the donor to fill its Roche lobe in a 50-minute orbit gives a radius of 0.082$\pm0.005$ \Rsun\ (twice as large as a cold WD of this mass).  

The assumption that mass transfer is driven only through gravitational radiation would then predict mass transfer rates below $10^{-10}$ \Msun/year \citep{Nelson86}, contrary to observations (Fig. 1), proving that additional angular momentum loss must be driving mass transfer.  
This information does not clearly discriminate between the He star and evolved main-sequence star evolutionary channels, as both can produce tracks roughly matching the orbital period, mass, and mass transfer rate of 4U 1916-053 (see Figs. 2 and 6 for the He star channel, and \citealt{Nelson03}'s track $M_c=0.0$, or \citealt{Podsiadlowski02}'s Fig. 15, for the latter channel).  Alternatively, donor wind mass loss or circumbinary disks could provide the required angular momentum loss in the WD scenario, but as discussed above we do not favor these possibilities.
 
4U 1916-053 is clearly below the irradiated helium disk stability line.  \citet{Lasota08} suggested that the presence of more than $\sim$5\% hydrogen in the disk would keep the disk stable and accretion persistent down to significantly lower mass transfer rates.  Would such a fraction be detectable, say, in studies of thermonuclear bursts?
A small hydrogen fraction $\sim10$\% would not affect the durations of bursts \citep{Cumming03}, nor would it be detectable in current optical spectroscopy \citep{Werner06,Nelemans10}.
However, due to hydrogen's larger energy release per nucleon, it could significantly change the energy released.  The energy released per nucleon is estimated at $Q_{nuc}=1.6+4.0<X>$ MeV/nucleon, where $<X>$ is the mean hydrogen fraction \citep{Cumming03,Fujimoto87}, which incorporates a 35\% energy loss to neutrino emission.  
\citet{Galloway08} measured the ratio of burst to persistent flux for 4U 1916-053, using a pair of bursts detected less than 10 hours apart with RXTE, as $\alpha=78.8\pm0.3$.  
Using \citet{Galloway08}'s relation between $\alpha$ and $Q_{nuc}$, 
\begin{equation}
\alpha=44 \frac{M}{1.4 \Msun}\left( \frac{R}{10 km} \right) ^{-1} \left( \frac{Q_{nuc}}{4.4 \mbox{MeV/nucleon}} \right) ^{-1}
\end{equation}
we find that a typical mass range of $M$=$1.4\pm0.2$ \Msun, $R$=11.5 km gives estimates of $<$$X$$>$=0.14$\pm.08$.  Thus, the energy release from burning hydrogen vs. helium  
suggests that 10-20\% of the accreted material should be hydrogen.  
This matches the predictions of a 10-20\% abundance of hydrogen at the surface of an evolved secondary star of period 49 minutes \citep{Nelemans10b,Nelson03}.  Such a fraction of hydrogen in the disk would nicely explain the persistence of this system. However, values of $\alpha$ may also be enhanced by incomplete burning of nuclear fuel, and often seem to vary with time in a single system, so further evidence of the existence of hydrogen should be sought.

A distinguishing characteristic between the He star and evolved main-sequence star tracks is that the He star tracks are evolving to longer periods, while the relevant evolved main-sequence star tracks are reaching their period minima at roughly this donor mass.  
\citet{Hu08} show that the orbit of 4U 1916-053 is expanding at the fast rate of $\dot{P}_{orb}/P_{orb}=1.62\times10^{-7}$ s$^{-1}$.  This is $\sim$100 times higher than the expected orbital period derivatives for the evolved main-sequence star scenario \citep{Nelson03}, and $\sim$5 times higher than the expected orbital period derivative for its mass and mass transfer rate, assuming conservative transfer where $\dot{P}_{orb}/P_{orb}=\frac{3 \dot{M}_d}{M_d} \left( \frac{M_d}{M_{accretor}}-1\right)$.

  Orbital period derivatives orders of magnitude larger than expected, and/or with the wrong sign, are a common problem of XRBs \citep[e.g.][]{Wolff09,Chou01b,Burderi09}. They can often be explained by the transfer of angular momentum between the donor and orbit on timescales of years, as seen in several X-ray binary or millisecond pulsar systems with very low-mass donors \citep{Arzoumanian94,Patruno12}.  However, it is unclear whether such an explanation can apply to partly degenerate donors such as 4U 1916-053.  The other option is nonconservative mass transfer, as predicted by the \citet{vanHaaften12} scenario.

\subsection{4U 1626-67}
The other persistent UCXB with a constraint on the donor mass is 4U 1626-67.  An upper limit derived from searching for timing variations in the X-ray pulses at the known 41.4 minute orbital period constrains the projected semimajor axis of the NS to $<$8 light-milliseconds \citep{Levine88,Shinoda90}.  This implies sin $i<7.8\times10^{-3} q^{-1} (1+q)^{2/3} M_{NS,1.4}^{-1/3} P_{42}^{-2/3}$  \citep{Chakrabarty98}, where $q$ is the donor/NS mass ratio, and $i$ is the inclination, measured from $i=0$ face-on.  \citet{Schulz01} found double-peaked emission lines in its spectrum, interpreted as the Keplerian Doppler shifts of the accretion disk.  \citet{Krauss07} measured disk line velocities of $v$ sin $i$=1700 km/s, which with the corotation radius of the NS at $6.5\times10^8$ cm \citep{Coburn02}, allows a constraint on the inclination angle, $i>$22 degrees.  Combining this with the projected semimajor axis upper limit constrains the donor mass to $<$0.036 \Msun\ for a 1.4 \Msun\ NS.  Requiring the donor to also fill its Roche lobe gives a radius of $<$0.06 \Rsun.

4U 1626-67 shows clear evidence of C, O, and Ne in its X-ray and optical spectra.  Thus, the evolved main-sequence star track is ruled out.  Its long-term accretion history has given a total fluence of 0.927 ergs cm$^2$ \citep{Krauss07}, or $>2.8\times10^{45}$ ergs for $d>$5 kpc \citep[the minimum distance derived considering optical reprocessing by][]{Chakrabarty98}.  For a 1.4 \Msun, 11.5 km NS, the inferred total mass transfer is $>1.7\times10^{25}$ g.  For comparison, the maximum mass of a cold quiescent helium disk in a 41.4 minute system is $\sim1.6\times10^{25}$ g \citep{Lasota08}.  The latter estimate depends on the poorly-known viscosity, and was calculated for helium rather than C/O.  Since C/O disks have lower instability limits (see \S 2.3), they will go into outburst at lower disk densities, and thus can store even less mass.   
Thus, it is very difficult to believe that 4U 1626-67 is undergoing a transient outburst; its mass transfer must indeed be persistent.  
This rules out the WD evolutionary track, which cannot evolve a WD donor from short periods \citep{Deloye03}.  

He star tracks can possibly explain the nature of 4U 1626-67.  Our He track with $\dot{M}=10^{-6}$ \Msun/year passes nicely through its orbital period and most likely mass transfer rate.  However, we can also calculate the mass of our simulated donors at 42 minutes, finding that the $\dot{M}=10^{-6}$ track gives a mass that is too high (Fig. 6).  In addition, the inferred density of the donor seems too high for the $\dot{M}=10^{-6}$ track (mass and radius limits above, and Fig. 3). 
The $\dot{M}=10^{-7}$ track is a more appropriate fit for both.  
If 4U 1626-67 is at the relatively nearby distance of $\sim$5 kpc (barely allowed within the distance errors; see Fig. 2, Table 1), its inferred luminosity and therefore $\dot{M}$ would also be consistent with the $\dot{M}=10^{-7}$ track.     Alternatively, the very high magnetic field of this NS ($3\times10^{12}$ G, \citealt{Coburn02}) may provide additional magnetic braking in this system, which we do not attempt to model here.  Finally, donor wind mass loss or circumbinary disks with a WD donor are also possibilities.

The past and future of 4U 1626-67 are both unusual.  To attain its current low-mass, high-entropy state, the donor star must have been fusing He when it started mass transfer, and thus must have transferred several tenths of a solar mass to the accretor.  Yet the accretor has a magnetic field of $3\times10^{12}$ G, compared to typical magnetic fields of $10^8$ G for recycled NSs, which have likely been driven to field decay due to accretion of a few tenths of a solar mass \citep{Bhattacharya91}.  It is difficult to understand why the magnetic field of the NS in 4U 1626-67 did not decay.  This suggests either that the NS in 4U 1626-67 underwent accretion-induced collapse \citep{Taam86,Yungelson02}, or that the progenitor NS had an originally much higher field strength, of magnetar levels.  
We cannot infer the age of this system from its spin period, since its changes from spin-up to spin-down \citep[e.g. ][]{CameroA10} indicate that it is close to spin equilibrium. 
  A difficulty with the accretion-induced collapse scenario is that mass transfer will stop after the collapse for a period of $\sim10^8$ years \citep{Verbunt90}, which should allow a low-mass donor to cool down and join the standard WD tracks \citep{Deloye07}.  As the analysis of \citet{Verbunt90} assumed rapid magnetic field decay in NSs,  a detailed reconsideration of this scenario could be rewarding.

\subsection{4U 0614+091}
The orbital period of 4U 0614+091 is not well-determined, although there are hints from both optical photometry and spectroscopy. \citet{Obrien05}, \citet{Shahbaz08}, \citet{Hakala11}, and \citet{Zhang12} have found photometric evidence for orbital periods of 50-51 minutes, although not consistently (for instance, Hakala et.~al.\ found a 50 minute periodic signal in only 1 of 12 datasets, and several of these papers found other possible periodicities). Weak evidence for a 48.5 minute orbital period was suggested by \citet{Nelemans06} from Gemini/GMOS spectroscopic data, but \citet{Madej13} find evidence for a 30 minute orbital period in the same Gemini/GMOS data as well as in VLT/X-Shooter spectra (and do not confirm the 48.5 minute period).  No convincing evidence of the orbital period is seen in X-ray observations \citep{Hakala11,Madej13}.

If the suggested 50-51 minute orbital period of this system is correct, then the persistent nature of this system and the C/O composition of the accretion disk \citep{Nelemans06,Schulz10,Madej10} suggest a He star donor, as for 4U 1626-67.  
A 30-minute orbital period would not change this conclusion, as the inferred mass transfer rate would still be rather larger than can be supplied by standard WD evolution.  However, the uncertainty in the orbital period means that conclusions about the nature of this object remain unclear.  
The clear evidence for He in the X-ray bursts, along with the lack of any evidence for He in optical spectra of this object, presents another mystery \citep{Kuulkers10}.



\section{Conclusions}

We utilize the luminosity histograms calculated in \citet{Cartwright12}, new calculations of He star evolution, and a wide range of key facts from the literature to place constraints on the nature of various individual UCXBs.  

We have calculated time-averaged mass transfer rates for UCXBs with known (or suggested) periods, and compared them with theoretical mass transfer rates.  Most agree very well with tracks for WD donors, supporting the idea that this is the primary route for UCXB production.  
A group of two or three systems with high ($>10^{-10}$ \Msun/year) mass transfer rates and long ($>$40-minute) periods require alternative explanations, as standard WD evolution is incapable of reaching them.  Circumbinary disks, while mathematically capable of providing the required angular momentum loss rates, require rather unfeasible physical conditions.  Enhanced donor wind mass loss rates (suggested by \citealt{vanHaaften12}) might explain these systems, but we identify some potential difficulties with this explanation.  We show that stellar cores burning helium when mass transfer starts (He stars) are capable of reaching these mass-transfer rates at these periods due to their much higher initial entropy and continued strong irradiation, and could explain all three systems.  

The disk instability line for helium accretion disks, even when irradiated by accretion X-rays, is too high to explain the persistent behavior of at least three, probably six, UCXB systems.  We point out that C/O disks have lower disk instability lines (as calculated by \citealt{Menou02}), which when considering irradiation can easily keep these systems persistent.  As four of the six show evidence of carbon and/or oxygen in their optical, UV, and/or X-ray spectroscopy, this nicely explains their behavior.  This provides a key physical explanation for the difference between the empirical UCXB luminosity function and current theoretical ones. 

One system, 2S 0918-549, lies well below the lowest-entropy helium WD track.  If its recent mass transfer rate is representative, it must have a C/O WD; this is consistent with its optical spectroscopy, which suggests a C/O disk.  A tentative 10.8 minute period for 4U 1728-34  would make this system's mass transfer rate impossible to explain by any evolutionary model, and require dramatic fluctuations in mass transfer on timescales $>$30 years.

Two unusual UCXBs have additional data to test evolutionary models.  4U 1916-053 shows positive and negative superhumps, due to its precessing accretion disk, which permit an estimate of the mass ratio and thus of the donor mass; we derive $M_{donor}$=0.064$\pm0.010$ \Msun (assuming a 1.4 \Msun\ NS).  4U 1916-053 shows helium and nitrogen in its optical spectra, which permits either an evolved main-sequence star or a slightly evolved He star as the initial donor.  The large orbital period derivative suggests a He star for the donor, but is not conclusive.

4U 1626-67 shows 7.7 s X-ray pulsations, but no pulse frequency shifts.  Combined with a constraint on the disk's  inclination from the velocity of X-ray lines in the disk, we can constrain $M_{donor}<$0.036 \Msun\ (for a 1.4 \Msun\ NS). 
 C, O, and Ne line emission has been observed in the optical, ultraviolet, and X-ray.   Its time-integrated X-ray luminosity indicates a total mass transferred onto the NS during its history of continuous accretion that exceeds the mass that can be stored in a C/O disk in a 42-minute orbit, proving that this system is not experiencing a transient disk-instability outburst, but is truly a persistent system.  The chemical composition and mass-transfer requirements strongly suggest a He star donor.  The unusually high magnetic field of the NS indicates a past history of substantially stronger, magnetar-strength fields or recent  accretion-induced collapse of a WD.

\acknowledgments

This research has made extensive use of ADS and the arXiv. We acknowledge financial support from NSERC (Discovery Grants to COH, NI, and GRS, and an NSERC USRA, Julie Payette NSERC Research Scholarship, and Andr\'e Hamer Postgraduate Prize supporting MCE), an Alberta Ingenuity New Faculty award to COH, a Canada Research Chair supporting NI, and the Avadh Bhatia Fellowship supporting JCG.  We thank the anonymous referee for a particularly thoughtful report.

\bibliography{src_ref_list}
\bibliographystyle{apj}

\end{document}